\documentclass{./aa}  

\usepackage{graphicx}
\usepackage{txfonts}
\usepackage{booktabs}
\usepackage{appendix}

\usepackage[nice]{units}
\usepackage{color}
\usepackage{natbib}

\begin{document}

	\title{LOFAR Measures the Hotspot Advance Speed of the High-Redshift Blazar S5\,0836+710}
	\titlerunning{LOFAR Measures the Hotspot Advance Speed of S5\,0836+710}

\author{A. Kappes\inst{1}, M. Perucho\inst{2,3}, M. Kadler\inst{1}, P. R. Burd\inst{1}, L. Vega-Garc\'ia\inst{4}, M. Brüggen\inst{5}
          }
\authorrunning{A. Kappes et. al}

   \institute{Institut für Theoretische Physik und Astrophysik, Universität Würzburg,
              Emil-Fischer-Straße 31, 97074 W\"urzburg, Germany
              \email{alexander.kappes@uni-wuerzburg.de}
            \and
            Departament d'Astronomia i Astrof\'isica, Universitat de Val\`encia, C/ Dr. Moliner, 50, E-46100 Burjassot, Val\`encia, Spain 
            \and
            Observatori Astron\`omic, Universitat de Val\`encia, C/ Catedr\`atic Beltr\'an 2, E-46091 Paterna , Val\`encia, Spain
           \and
              Max-Planck-Institut f\"ur Radioastronomie,
              Auf dem H\"ugel 69, D-53121 Bonn, Germany
              \and
              Hamburger Sternwarte, Universität Hamburg, Gojenbergsweg 112, 21029 Hamburg, Germany
             }

   \date{Received TBD; accepted TBD}

  \abstract
{The emission and proper motion of the terminal hotspots of AGN jets can be used as a powerful probe of 
    the intergalactic medium.
	However,
	measurements of hotspot-advance speeds in active galaxies are difficult, 
	especially in the young universe,
	due to the low angular velocities and the low brightness of distant radio galaxies.
}
{Our goal is to study the termination of an AGN jet in the young universe and to deduce physical parameters of the jet and the intergalactic medium.}
{We use the LOw Frequency ARray (LOFAR) to image the long-wavelength radio emission of the high-redshift blazar S5\,0836+710 on arcsecond scales between 120\,MHz and 160\,MHz.}
{The LOFAR image shows a compact unresolved core and a resolved emission region  about 1.5\,arcsec  to  the  southwest  of  the  radio  core. This structure is in general agreement with previous higher-frequency radio observations with MERLIN and the VLA. The southern component shows a moderately steep spectrum with a spectral index of about $\gtrsim -1$ while the spectral index of the core is flat to slightly inverted. In addition, we detect for the first time a
resolved steep-spectrum halo with a spectral index of about $-1$ surrounding the core. }
{The arcsecond-scale radio structure of S5\,0836+710 can be understood as an FR\,II-like radio galaxy observed at a small viewing angle.  The southern  component  can  be  interpreted as  the region of  the  approaching  jet's terminal hotspot and  the  halo-like  diffuse  component near the core can be interpreted as the counter-hotspot region. From the differential Doppler boosting of both features, we can derive the hotspot advance speed to $(0.01-0.036)$\,c. At a constant advance speed, the derived age of the source would exceed the total lifetime of such a powerful FR\,II-like radio galaxy substantially. Thus,  the  hotspot advance  speed must have been  higher in the past in agreement with a scenario in which the originally highly relativistic jet has lost collimation due to the growth of instabilities and has transformed into  an  only  mildly  relativistic  flow.
Our data suggest that the density of the intergalactic medium around this distant ($z=2.22$) AGN could be substantially higher than the values typically found in less distant FR\,II radio galaxies. 
}

\maketitle
%
\section{Introduction}
\label{sec:intro}
Radio-loud active galactic nuclei (AGN) can eject powerful double-sided relativistic jets \citep{blandford2018} into the intracluster medium that emit synchrotron emission and can reach distances of Megaparsecs.
The most powerful of these sources exhibit so-called FR\,II morphologies \citep{fanaroff1974} in which the jets are terminated in high surface-brightness regions called hotspots, where the jets interact with the surrounding medium. FR\,II radio galaxies have been studied extensively at centimeter wavelength with the Very Large Array \citep[VLA; e.g.,][]{odea2009}, estimating ages, velocities, magnetic fields, total lifetime, ambient gas densities, and other quantities.

Blazars are a subclass of AGN, whose jets are aligned at a small angle to the line of sight towards Earth. They can be classified into BL\,Lac objects and flat-spectrum radio quasars (FSRQs). According to
the AGN unified scheme \citep{antonucci1993,urry1995}, FSRQs are the beamed counterparts of FR\,II radio galaxies. Because of relativistic bulk motion of plasma at small inclination angles, the compact (i.e., parsec scale) emission of blazar jets gets drastically Doppler boosted and can be observed out to very high redshifts.

According to synchrotron theory, electrons with Lorentz factor $\gamma$ emit at 
frequencies $\nu \sim 10^{-6} \gamma^2 B^2$\,GHz (with $B$ in mG). High radio-frequency
observations thus typically probe emission from electrons at $\gamma > 1000$.
The emitted spectrum of large-scale components associated with AGN jets (unbeamed lobe
emission and moderately beamed hotspot emission) can be described by a power law, $F_\nu\propto \nu^{\alpha}$, with the spectral index $\alpha$ typically in the range $-0.5$ to $-1)$, while the beamed emission from the central jet in blazars  is typically characterized by
flat spectral indices $\alpha \sim 0$. Consequently, blazars have been studied extensively
at high radio frequencies, where the jet dominates and Very-Long-Baseline Interferometry (VLBI) techniques offer unprecedented
angular resolution of the inner jet region \citep{zensus1997}, while the low-frequency properties
of blazar lobe-emission have received less attention in the past decades.

Several further constraints affect the study of large-scale blazar observational data. Due to the strong projection effects at the small inclination angles involved, the emission of the two lobes associated with the jet and counterjet can blend. Moreover, the hotspots of jet and counterjet are subject to noticable differential Doppler boosting due to the mildly relativistic advance speeds of hotspots in the intergalactic medium \citep{odea2009}. Light-travel time differences between both jets can affect the observed arm ratios and cause differential
aging of hotspot and counter-hotspot. 

These problems can partially be overcome with high-resolution observations at long observing wavelengths 
as provided by LOFAR \citep{van2013}, which offers unprecedented sensitivity at $40 - 240$\,MHz 
and an angular resolution greatly improved
over previous instruments.
The low-energy electron population ($\gamma < 1000$) responsible for
the blazar-lobe emission can give us unique new insights into the large-scale structure
of blazars and therefore a probe of the oldest observable structures in these powerful
sources. In particular, while the counterjets of blazars are typically strongly debeamed and therefore
unobservable, the lobe and hotspot emission associated with these counterjets are expected to be less
strongly debeamed. Due to their steep radio spectrum and small projected scales, they can be detectable in LOFAR observations, while having remained undetected in previous higher-frequency and/or lower angular-resolution observations.

The powerful high-redshift ($z=2.218$) blazar S5\,0836+710 
has been observed with the VLA  by \citet{cooper2007} at 1.4\,GHz. At this moderately low frequency, the VLA (in A configuration) did not resolve the kilo-parsec-scale structure of the source. Higher-frequency VLA observations \citep[e.g.,][]{odea1988} and observations with MERLIN at 1.6\,GHz \citep{hummel1992} have shown a single extended and polarized emission feature about 1.5\,arcsec south of the jet core without any visible emission bridge between it and the core and without any apparent counterpart on the other side of the core.
\citet{perucho2012a} and \citet{perucho2012b} suggested that the jet in S5 0836+710 is subject to the development of Kelvin-Helmholtz (KH) instability \citep{perucho2012a} and that this instability could be the cause of jet disruption and the generation of a decollimated radio structure at arcsecond scales \citep{perucho2012b}, explaining the prominent extended feature observed by \citet{odea1988} and \citet{hummel1992}. However, \citet{perucho2012b} pointed out that the disruption site should be associated with intense dissipation of kinetic energy, which is not observed at any point between the inner jet and the putative relic feature. Another problem in the jet-disruption scenario is related to the one-sided kiloparsec-scale morphology because no corresponding relic or lobe associated with the counterjet can be observed. In this paper we present new results from LOFAR observations that solve these problems. The overall extended arcsecond-scale structure can be interpreted as a classical but strongly projected double-sided source morphology in which the southern feature is a hotspot associated with the approaching jet rather than a disrupted jet relic.

The following sections are structured as follows: In Sect.~\ref{sec:data_observation} we give observational parameters of the LOFAR observation of S5\,0836+710 and describe the data reduction. The resulting images and derived quantities are presented in Sect.~\ref{sec:results}. Section~\ref{sec:discussion} presents a discussion of the observational data and implications. Throughout the paper, we use the following cosmological parameters: $H_0=71$\,\nicefrac{km}{s$\cdot$Mpc}, $\Omega_\text{m}=0.27$ and $\Omega_\Lambda=0.73$.

\section{Observation and data reduction}
\label{sec:data_observation}

\begin{figure}[h]
	\centering
	\includegraphics[width=\linewidth]{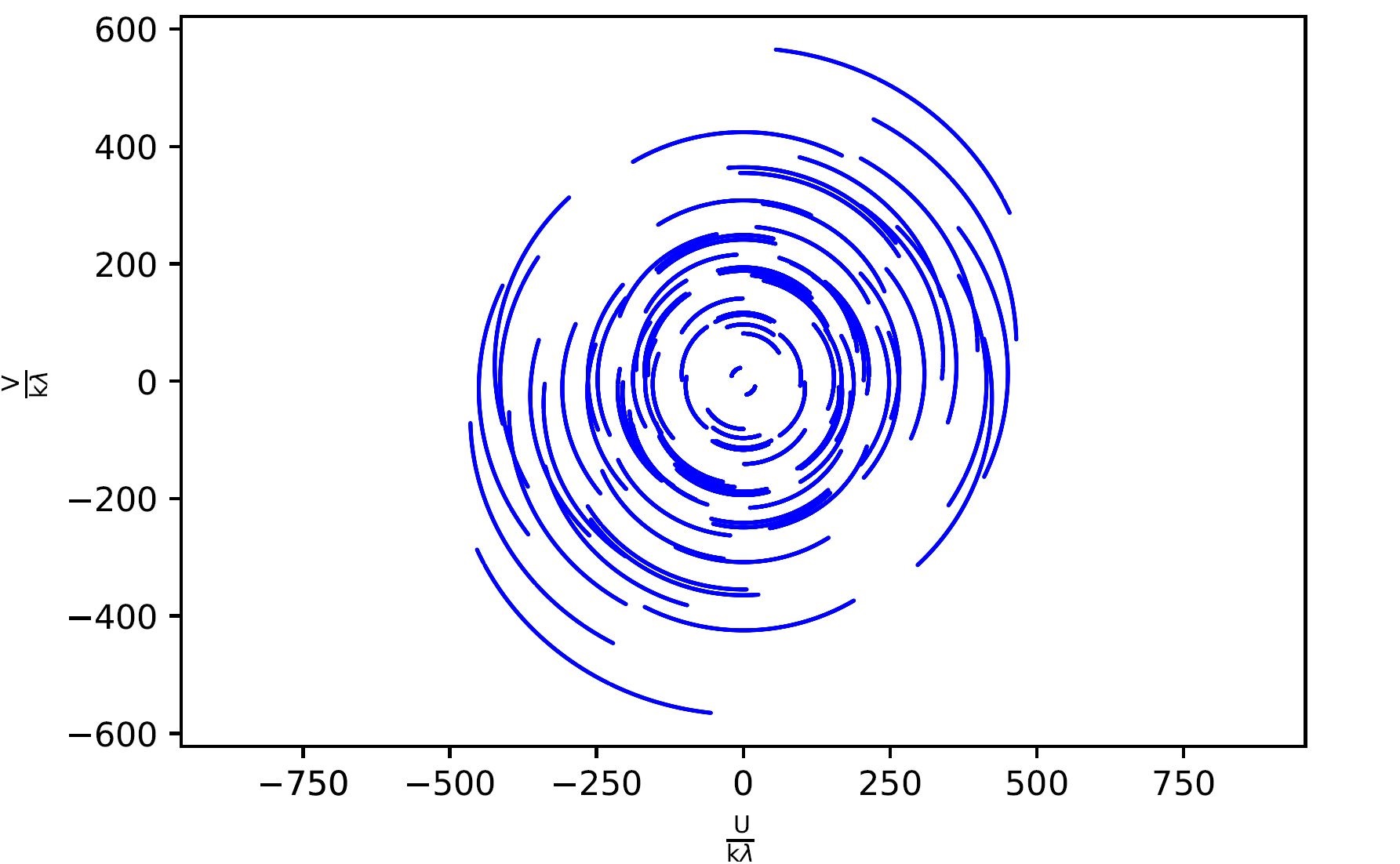}
	\caption{(uv)-coverage of the international-LOFAR data set at 135 MHz. 
	}
	\label{fig:uvplane}
\end{figure}

\begin{figure}[h]
	\centering
	\includegraphics[width=\linewidth]{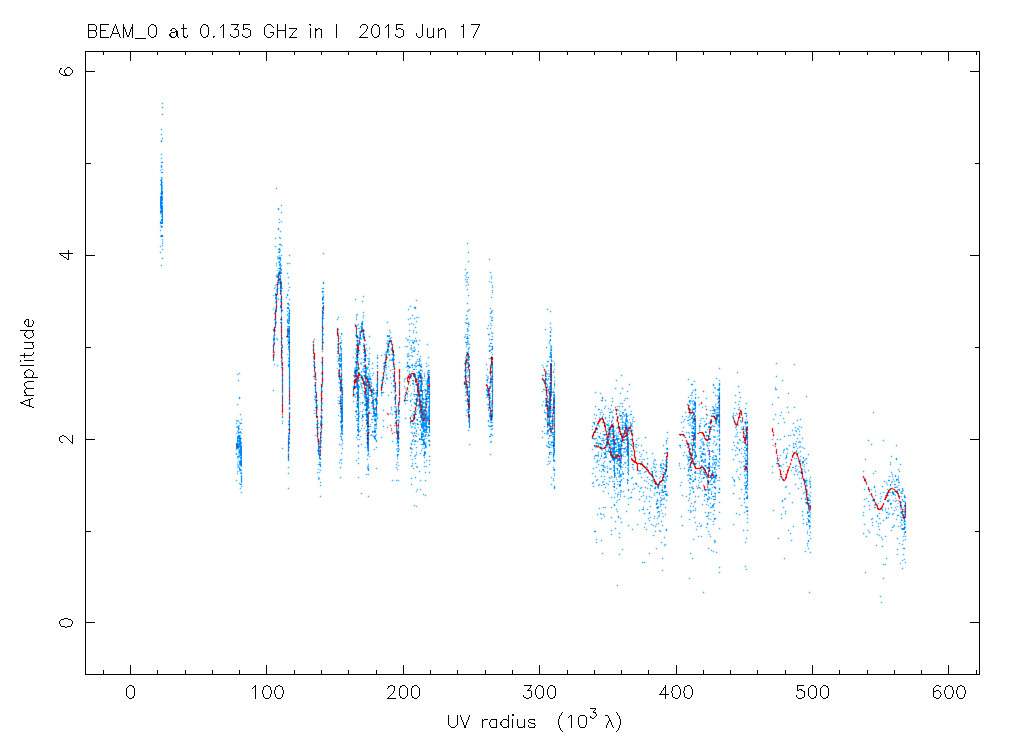}
	\caption{Measured visibilities in the 135\,MHz LOFAR data set as a function of (u,v)-radius. Blue dots represent visibilities, while red dots show the model developed in the hybrid imaging process.
	}
	\label{fig:radplot}
\end{figure}

We observed S5 0836+710 on June 17, 2015, with the LOFAR High Band Antenna (HBA) array. The observation time was 4 hours covering the full frequency range between 117.5\,MHz to 162.6\,MHz. 3C\,196, which was used as the primary flux-density calibrator,  was observed for 10 minutes at the beginning of the observing run. Data were recorded in 8-bit mode in 
231 subbands with a bandwidth of 192 kHz each and averaged into
14 frequency bands, each with 3.12\,MHz bandwidth, and correlated with the COBALT (COrrelator and Beamforming Application platform for the Lofar Telescope) correlator \citep{broekama2018}. 
Given the known highly compact structure of S5\,0836+710, we used only the international LOFAR stations\footnote{In June 2015 this included 6 stations in Germany (DE601-DE605,DE609), and one station in Sweden (SE607), France (FR606) and the United Kingdom (UK608), each.} and analyzed the data using standard methods of Very-Long-Baseline Interferometry \citep[e.g.,][]{moran1995} using the AIPS \citep[Astronomical Image Processing System;][]{greisen2003} package. In this process, the data were averaged over 16 seconds in time so that the field of view contained only the target source.
The resulting (uv)-coverage is shown in Fig.\,\ref{fig:uvplane} and the measured visibilities as a function of (u,v)-radius are shown in Fig.\,\ref{fig:radplot}. Images were created using self-calibration techniques using \textsc{difmap} \citep{shepherd1997} directly on target, which was possible because of the bright and compact core emission present in S5\,0836+710. Four frequency bands had to be discarded due to insufficient data quality, presumably due to the bandpass shape and radio-frequency interference (RFI).
The total flux density picked up in the full image was calculated and used to compute a correction factor by comparing to the total flux density of the flux calibrator known from a low-resolution image made with the calibrated LOFAR core stations. This factor was applied to the high-resolution model and the visibilities between the international stations were self-calibrated with this corrected model. This process led to fully calibrated LOFAR VLBI images.

\section{Results}
\label{sec:results}
\begin{figure}[h]
	\centering
	\includegraphics[width=\linewidth]{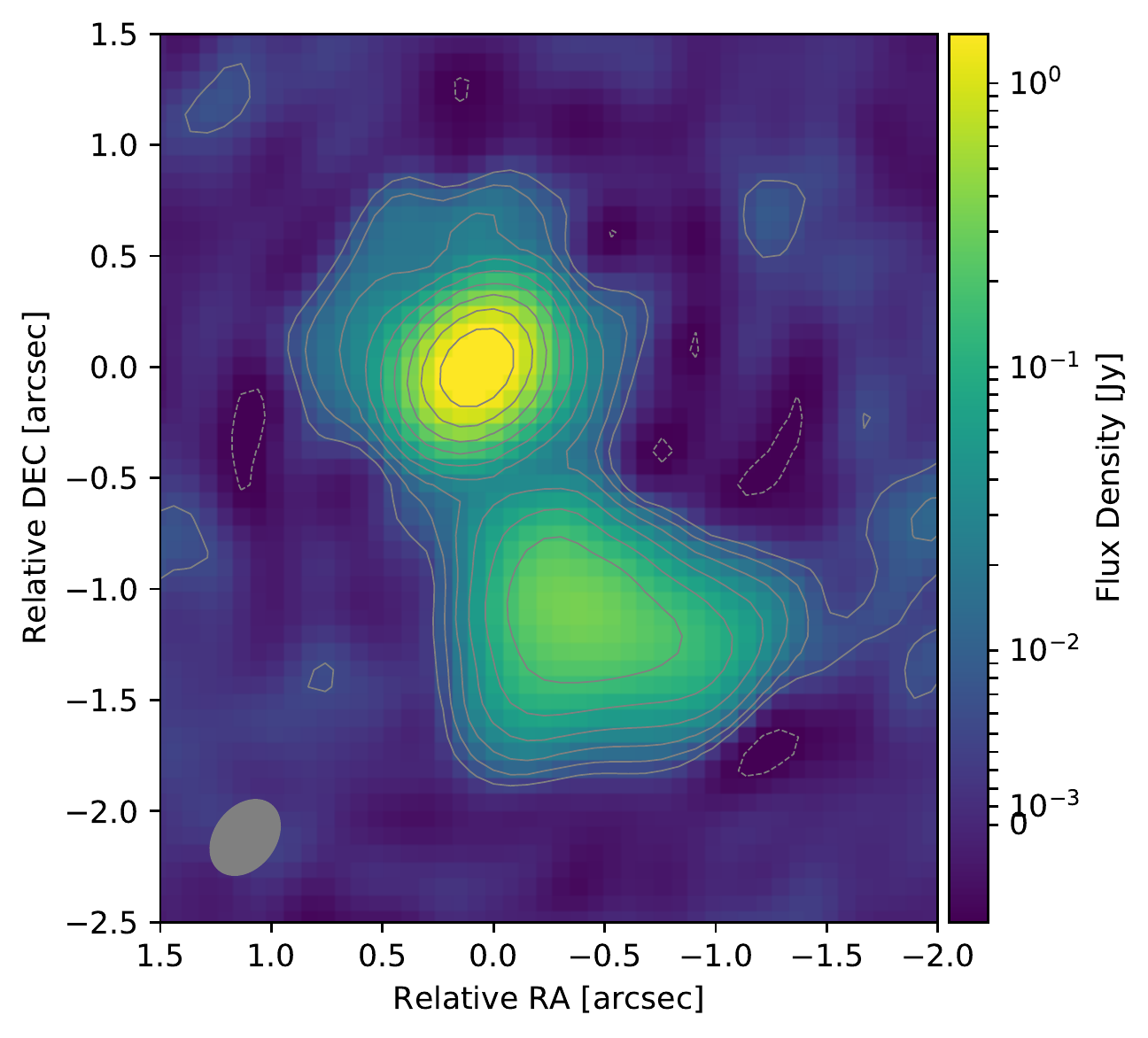}
	\caption{LOFAR stacked image of eleven frequency bands resulting in an effective central frequency of 143\,MHz and a 34\,MHz bandwidth. The lowest contour level corresponds to 3$\sigma$ significance. The contour levels are drawn at ($-$2, 2, 4, 8, ... ) times 2.7\,mJy\,beam$^{-1}$. The RMS is 3.1\,mJy\,beam$^{-1}$. The beam size (shown in the bottom left) is $0.459^{\prime\prime}\times0.308^{\prime\prime}$ with a position angle of $-39.1^\circ$. See online material for images of the individual-bands.}
	\label{fig:lofarstack}
\end{figure}

\,\ref{fig:lofarstack} shows a stacked image of the 11 bands (see online material for the images of the individual bands) corresponding to a central frequency of 143\,MHz and a bandwidth of 34\,MHz.

The general structure is in agreement with previous higher-frequency observations of S5\,0836+710 on comparable scales and angular resolution {\citep[see especially Fig.\,5 and Fig.\,1 in][respectively]{hummel1992,perucho2012b}}. {The source} shows a compact unresolved core and a resolved emission region between 1 and 2 arcsec to the southwest of the radio core. 
The core is known to contain a southward-directed compact VLBI jet with an extent of about 200\,mas or 1.5\,kpc \citep{perucho2012a}, that shows signs of growing instabilities with distance downstream. These were thought to lead to a full disruption of the jet before it is able to reach arcsecond scales \citep{perucho2012b}. In this scenario, the southern component was interpreted as a subrelativistic relic of the disrupted jet that continues propagating downstream and interacting with the intergalactic medium. Such features are generally expected to show steep spectral indices\footnote{We use notation $S_\nu \propto \nu^\alpha$, in which a negative spectral index $\alpha$ corresponds to a flux density $S_\nu$ falling with frequency.} with $-2 < \alpha < -1$ \citep{pandey-pommier2016}.

\begin{figure}[h]
	\centering
	\includegraphics[width=\linewidth]{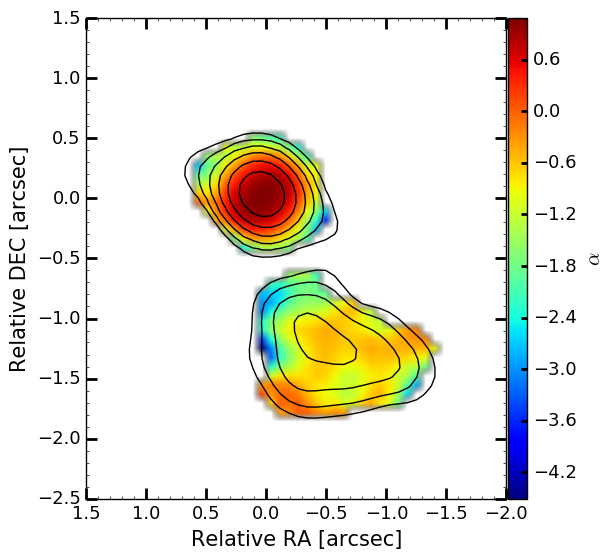}
	\caption{Spectral index map of eleven frequency bands model-fitted pixel-wise. The contour levels are obtained from the 129\,MHz image drawn at ($-$2, 2, 4, 8, ... ) times 15~\,mJy\,beam$^{-1}$. See online material for images of the individual-bands from which this map has been produced.}
	\label{fig:lofarspix}
\end{figure}

To test this scenario, we created a spectral-index image using the 11 individual frequency bands and fitting a power-law to each individual pixel (see Fig.~\ref{fig:lofarspix}). The southern component shows a moderately steep spectrum with a spectral index of about $\gtrsim -1$ while the spectral index of the core is flat to slightly inverted. An additional striking feature of the LOFAR spectral-index image is a resolved steep-spectrum halo with a spectral index of about -1 surrounding the core. This halo has not been seen in previous higher-frequency images of S5\,0836+710 and is only revealed by the good sensitivity and high angular resolution of LOFAR in the sub-GHz regime.

\begin{figure}[h]
	\centering
	\includegraphics[width=\linewidth]{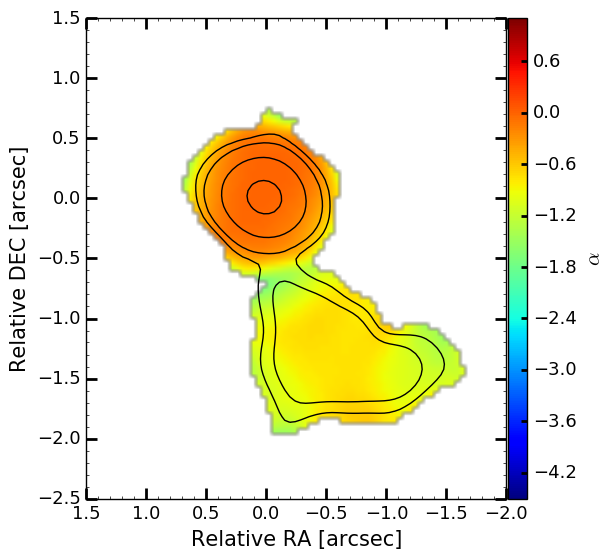}
	\caption{Spectral index image between 1.6\,GHz (MERLIN) and 138\,MHz (LOFAR). The contour levels are obtained from the MERLIN image and are drawn at ($-$2, 2, 4, 8, ... ) times 30\,mJy\,beam$^{-1}$.}
	\label{fig:lofarmerlinspix}
\end{figure}

~\newline
For comparison, we have produced a second spectral index map between the LOFAR band at 138\,MHz and 1.6\,GHz \citep[obtained from the MERLIN observation on \mbox{March 1, 2008}; see][]{hummel1992}, which is shown in Fig.~\ref{fig:lofarmerlinspix}. The core-spectral index is affected by source variability between the LOFAR (2015) and the MERLIN (2008) observation but shows a roughly flat spectrum. The steep-spectrum halo is marginally visible and its spectral slope is consistent with the LOFAR-only spectral-index image (Fig.~\ref{fig:lofarspix}). The southern component shows a spectral index of about $-0.7$ between 138\,MHz and 1.6\,GHz and is therefore also consistent with the LOFAR-only data within the given accuracy.

~\newline

\begin{figure}[h]
	\centering
	\includegraphics[width=\linewidth]{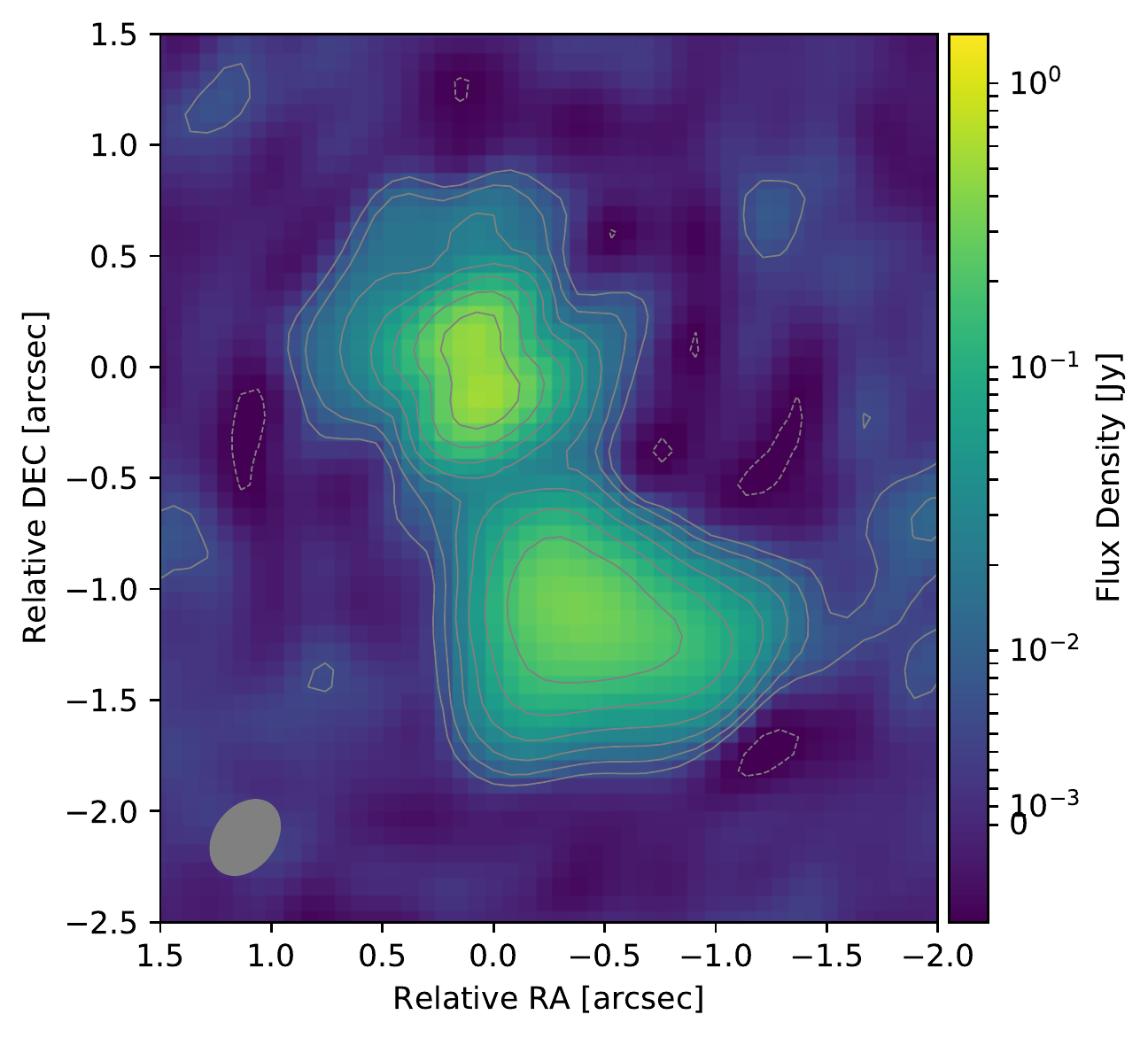}
	\caption{Residual LOFAR image after subtraction of the 2D-Gaussian modeled core component. The lowest contour is at 3$\sigma$ significance. Contour levels are shown at ($-$2, 2, 4, 8, ... ) times 2.7\,{mJy}\,{beam}$^{-1}$. The RMS is 3.1\,mJy\,beam$^{-1}$. The beam size (shown in the bottom left) is $0.459^{\prime\prime}\times0.308^{\prime\prime}$ with a position angle of $-39.1^\circ$.}
	\label{fig:nocore}
\end{figure}

Both spectral-index images consistently suggest that LOFAR resolves the central emission region into a core-halo structure. To test this, we model-fitted the LOFAR visibility data in this region with two superimposed Gaussian components in the image domain. We represented the core with an unresolved bright Gaussian component and modeled the halo with a broader and fainter Gaussian component. For visualization, we subtracted the unresolved core component from the full LOFAR image to obtain the brightness distribution shown in Fig.~\ref{fig:nocore}. The residual image reveals the  diffuse emission centered at a small offset of $\lesssim 0.01$\,arcsec from the core. 
The size of this second diffuse component is approximately the same as the size of the southern component and $(1.2 \pm 0.2)$\,Jy in flux density, translating to an intrinsic luminosity of \mbox{$(4.1\pm0.7) \cdot 10^{29}$\,{W}\,{Hz$^{-1}$}}.

\section{Discussion}
\label{sec:discussion}
In this section, we provide circumstantial evidence which shows that the large-scale morphology of S5\,0836+710 can be understood as an {classical} FR\,II-like radio galaxy seen at a small inclination angle. In this model, the southern
component can be understood as a face-on hotspot, hotspot relic, or hotspot-lobe structure of the approaching jet and the halo can be associated with emission of the hotspot and/or lobe on the counter jet side. We use the observational parameters to derive jet parameters and constrain the density of the intracluster medium surrounding the radio source S5\,0836+710.

\subsection{Interpretation of the southern emission region as a face-on hotspot} 
The spectral index 
of the southern component 
in S5\,0836+710 is only moderately steep and
does not reach values expected from  
a bona-fide radio relic of a disrupted jet as seen in other sources \citep{pandey-pommier2016}. The magnetic field in this region is circumferential at the south-eastern edge \citep{odea1988} as it is typical for quasar hotspots \citep{Swarup1984}. The field seems to be aligned with the western region of the large-scale structure. Altogether, it can be interpreted in this respect, as an emission region associated to a hotspot plus a lobe, in which the field is aligned along the shock in the west, as observed in other FRII sources \citep{Kharb2008}.
We therefore measured the size of the southern component by model fitting a Gaussian component to the LOFAR data in all 11 bands to test whether its extent is consistent with an active FR\,II hotspot region seen face on. The full extent of the emission region is about 9\,kpc with an average flux density of $(1.4\pm0.3)$\,Jy, translating to an intrinsic luminosity of \mbox{$(5\pm1) \times 10^{29}$\,{W}\,{Hz}$^{-1}$}. At a distance of 17.88\,Mpc, this is indeed on a scale of typical hotspot diameters in powerful FR\,II radio galaxies \citep{jeyakumar2000,perucho2003,kawakatu2006}. 

Alternatively, it is possible to find a model representation where only the brightest peak of the southern component is modeled with a Gaussian of about 5\,kpc in diameter and a flux density of $(1.2\pm0.2)$\,Jy (while residual emission on somewhat larger scales can be represented either by an additional wider Gaussian or by a hybrid model invoking a distribution of CLEAN components).
This latter representation would model a physical scenario of a hotspot surrounded by a lobe. Also in this model representation, the size of the high surface-brightness feature is still consistent with typical sizes of hotspots in FR\,II radio galaxies. 

The 'unusual' irregular morphology of the putative hotspot might indeed just be an effect of the high angular resolution and the small inclination angle at which the system is observed. If,
as suggested by \citet{perucho2012b}, this southern emission component does represent the relic of the hotspot, after the jet has been transformed into a subrelativistic or mildly relativistic broad flow, then the loss of collimation must have taken place fairly close upstream of the terminal feature, because it obviously has not expanded substantially since then.

\subsection{Interpretation of the source morphology as an FR\,II-like radio galaxy at a small viewing angle}  
\begin{figure}[h]
	\centering
	\includegraphics[width=\linewidth]{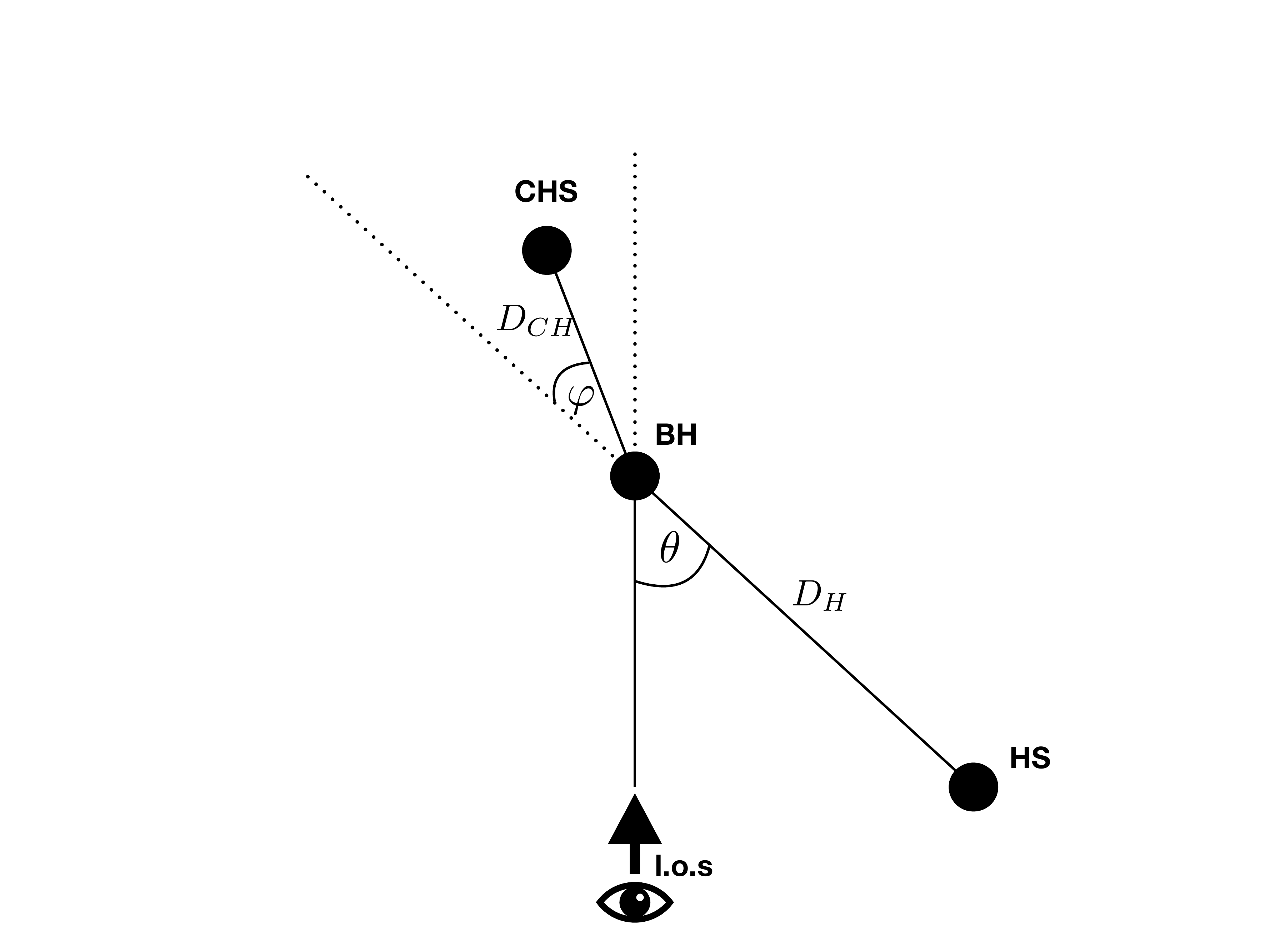}
	\caption{Model of the source geometry in S5\,0836+710 as an FR\,II radio galaxy observed at a small inclination angle $\theta$. HS denotes the {hotspot region} closer to the observer, and CHS denotes the {counter-hotspot region}. The different arm lengths are due to the different light travel times (see Appendix~\ref{sec:appendix_A}). The two jets are misaligned by $\varphi$ from a straight jet/counter-jet axis.}
	\label{fig:sketch}
\end{figure}

The kiloparsec-scale structure of S5\,0836+710 is consistent with a double-sided source, reminiscent of a highly projected radio-galaxy image onto which a strongly beamed unresolved core component is superimposed. In this interpretation, the southern diffuse component can be interpreted to be associated with the hotspot region of the approaching jet and the halo-like diffuse component near the core can be interpreted as the {counter-hotspot region}. Because the distance to the core is larger for the hotspot than it is for the counter-hotspot, the system cannot be fully symmetric. However, at small inclination angles, intrinsically small bends or misalignment angles can be increased to substantially larger apparent offsets in projection. A possible geometry of the system is shown in Fig.~\ref{fig:sketch}.

Interpreting the southern emission feature as the hotspot region and the halo-component near the core as the {counter-hotspot region} we calculate the brightness ratio of the two regions to be \mbox{$F_\textrm{h}/F_\textrm{ch}=(1.19\pm0.11)$}. 
This brightness ratio can be used to constrain the parameter space for inclination angle, misalignment and the advance speed $\beta_\textrm{h}$ of the jet head, which we assume to be the same for both hotspots. Advance speeds in FR\,II radio galaxies are generally assumed to be mildly relativistic with values of up to $0.1$c to $0.5$c \citep{odea2009}, depending on the deviation from the minimum-energy conditions. Under a small inclination angle, such speeds can lead to notable differential boosting effects.
In this framework, we thus expect

\begin{equation}
\label{eq:dopplerad}
\dfrac{F_{\mathrm{h}}}{F_{\mathrm{ch}}}=\left(\dfrac{1+\beta_\textrm{h} \cos\theta}{1-\beta_\textrm{h} \cos{\left(\theta-\varphi\right)}}\right)^{3-\alpha}\quad,
\end{equation}
where $\theta$ is the inclination angle under which we observe the approaching jet and $\varphi$ is the misalignment angle of the counterjet with respect to the approaching jet (see Fig.~\ref{fig:sketch}). 

\begin{figure}[h]
	\centering
	\includegraphics[width=\linewidth]{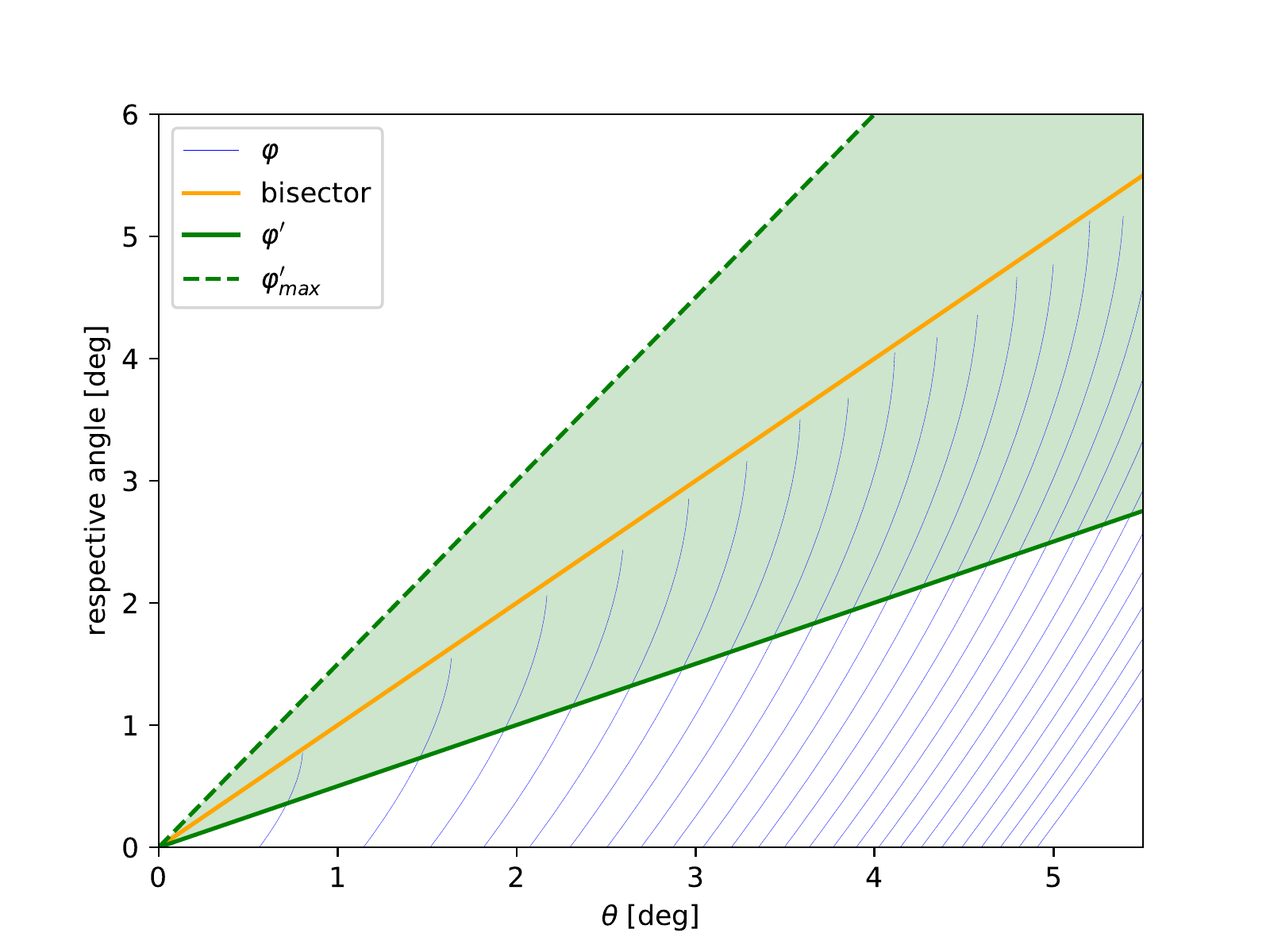}
	\caption{$\varphi$ lines show the misalignment needed to explain the measured flux ratio with respect to a certain $\beta$. $\beta$ values are represented by the drawn arcs, and increase from 0.02339 (leftmost arc) up to 0.02348 (rightmost arc) in increments of \mbox{$3.4\cdot 10^{-6}$}. $\varphi'$ is the minimum needed, and $\varphi'_{max}$ is the maximum possible misalignment angle resulting from the geometric argument (see text). The confined green area in between is therefore the possible misalignment in order to blend the {counter-hotspot region} by the bright core.}
	\label{fig:paramspace}
\end{figure}

With the measured spectral index of $\alpha=-0.7$, the measured brightness ratio of the two hotspot regions, and a viewing angle of $\theta = 3.2^\circ$, as estimated by \cite{pushkarev2009jet}, this relation constrains the allowed parameter space as seen in Fig.~\ref{fig:paramspace}.
For a given flux ratio, the resulting head advance speed
does depend only weakly on the geometry,
Considering the uncertainty range of the flux-ratio, the advance speed is constrained to the range of 0.010\,c to 0.036\,c. This is comparable to (albeit on the low end of the distribution of) source advance velocities of distant high-power FR\,II radio galaxies \citep{odea2009}.

Hotspot advance speeds of active FR\,II radio galaxies have been commonly assumed to be roughly constant over the lifetime of a source \citep{odea2009}. In that case, the measured advance speed for S5\,0836+710 would imply a source age of $2 \times 10^7$ years to $8 \times 10^8$ years, which is exceeding the maximum total source lifetime of such a powerful source by a factor of $2$ to $80$ \citep[e.g.,][]{odea2009,perucho2019}. Thus, the hotspot advance speed in S5\,0836+710 must have been somewhat higher in the past, in agreement with a scenario in which the originally highly relativistic jet has lost collimation due to the growth of instabilities and has transformed into an only mildly relativistic flow, as suggested by \citet{perucho2012b}.

The hotspot region was modeled with a single Gaussian component for each band. Averaging all bands, we measure an  apparent opening angle (core to hotspot) of $(25\pm 2)^\circ$. From that, it can be derived that the
inclination angle is unlikely to be much larger than $15^\circ$, because that would imply an intrinsic opening angle of $\gtrsim 7^\circ$.
On the other hand, it is highly improbable that the inclination angle is much smaller than about $1.5^\circ$ because otherwise the total deprojected source size of $>1$\,Mpc would be larger than the maximum known sizes of radio sources \citep{jeyakumar2000}. At the preferred inclination angle of $\theta = 3.2^\circ$ \citep{pushkarev2009jet}, the measured apparent opening angle implies an intrinsic opening angle of about $1^\circ$, which is consistent with the conclusions of \cite{hummel1992}.

An independent additional constraint on the inclination and the misalignment angles follows from the simple geometric argument that we see the {hotspot region} at a distance of about 1.5\,arcsec from the core and the {counter-hotspot region} is located within about $0.5$\,arcsec from the core (because at that distance the core and the counter-hotspot start to merge into a blended feature at the beamsize of our LOFAR image). We thus require a misalignment angle $\varphi \gtrsim 2.5^\circ$.

The misalignment $\varphi$ needed to explain both the observed morphology and the hotspot regions brightness ratio is thus in the range $2.5^\circ$ to $5^\circ$. {Within the range of typical inclinations of blazars towards earth, this leaves a relatively large parameter space (see Fig.~\ref{fig:paramspace}), that is fully consistent with expectations for the special case of S5\,0836+714.} Such values are indeed common among powerful FR\,II radio galaxies. For example, misalignment angles in 3C sources  are known to range up to values of about $12^\circ$ or more \citep{leahy1984}.
We thus consider the interpretation of the 
S5\,0836+710
large-scale morphology as an FR\,II-like radio galaxy at a small viewing angle as realistic.

\subsection{Derivation of the jet parameters}
\label{sec:jet_param_deriv}

The total jet power for a relativistic jet is parametrized as \citep[see, e.g.][]{perucho2017}:

\begin{equation}\label{eq:ljr}
L_{\rm j}\,=\,\left( \rho_{\rm j} h_{\rm j} \Gamma_{\rm j}^2 \,+\,\frac{(B^{\phi})^2}{4\pi}\right) v_{\rm j}\,A_{\rm j}, 
\end{equation}

where 

$ h_{\rm j} = c^2 + \frac{\gamma_{\rm j} P_{j}}{(\gamma_{\rm j} -1)\rho_{\rm j}}$

is the jet specific enthalpy, ${\bf \Gamma}_{\rm j}$ is the jet Lorentz factor, $\rho_{\rm j}$ is the jet rest mass density, $v_{\rm j}$ is the jet velocity, $c$ is the speed of light, $B^\phi$ is the toroidal field in the observer's frame, $\gamma_{\rm j}$ is the ratio of specific heats of the jet gas, and $A_j$ is the jet cross-section. The first term in Eq.~\ref{eq:ljr} includes the kinetic, internal, and rest-mass energy contributions, while the second term stands for the magnetic energy of the jet.

We can assume (see Appendix~\ref{sec:appendix_B}) that the jet is kinetically dominated and in the cold regime (i.e., its magnetosonic Mach number is high). Given also the fact that the advance speed as measured by LOFAR is so slow, the velocity of the bulk plasma flowing into the hotspot can be assumed to be only mildly relativistic at most (see also Appendix~\ref{sec:appendix_C} for the relativistic derivation). Under these conditions, the magnetic and pressure terms can be neglected and Eq.~\ref{eq:ljr} simplifies to

\begin{equation}\label{eq:ljca}
L_{\rm j}\,=\, \frac{1}{2}\,  v_{\rm j}^3\, \rho_{\rm j}\, A_{\rm j}.  
\end{equation}

Across a strong shock, the hotspot pressure is \citep[e.g.,][]{ll59}

\begin{equation} \label{eq:ph}
P_{\rm h} \simeq \frac{4\,L_{\rm j}}{(\gamma_{\rm j} + 1) v_{\rm j} A_{\rm j}}.
\end{equation}

In this equation, we need to define the jet radius at the hotspot, $R_{\rm j,h}$, which can be approximated as the hotspot radius. Because we do not know whether the southern radio-structure includes the hot-spot and part of the lobe, or it is the hot-spot, we will consider both half of the whole region size as the jet radius ($4.5\,{\rm kpc}$), and half the size of the fitted component, which represents the brightest region within the hypothetical lobe ($2.5\,{\rm kpc}$). Although the polarization seems to favour the latter interpretation (see the previous section), we study both cases here. Furthermore, we need an estimate for $v_{\rm j}$ at the hotspot, $v_{\rm j, h}$. As we will show in the following paragraphs, any reasonable input value is sufficient because we define an iterative method to derive its value of convergence.
Once the hotspot pressure is obtained, we can use equipartition between the non-thermal particles and the magnetic field, as reported for FR\,II hotspots \citep[see e.g.,][]{hardcastle2000}, to obtain a value of the magnetic field at the interaction site.  

The value of the magnetic field prior to the reverse shock $B^{\phi}_{\rm{j,h}}$ can be constrained by assuming conservation of the magnetic flux from the 1.6\,GHz jet to the interaction site\footnote{This assumption is quite reasonable for conically expanding jets at kiloparsec scales \citep[see e.g.,][]{komissarov2012book}.}. Applying the MHD jump conditions at the reverse shock giving rise to the hotspot:
\begin{equation}
v_{\rm j, h} B^{\phi}_{\rm j, h} \, = \, v_{\rm h} B^{\phi}_{\rm h},
\end{equation}   
allows us to derive a new estimate for $v_{\rm j, h}$. Setting a convergence criterion for this parameter at a precision of $10^{-3}$, we can find the relevant parameters of the problem. 

We have applied this method to four different sets of hotspot velocity and radius, namely, the possible combinations of $v_{\rm h}\,=\, 0.01 - 0.036\,c$, and $R_{\rm j, h}\,=\,2.5 - 4.5 \,{\rm kpc}$. Table~\ref{tab:parsh} shows the resulting values. The ranges given for the parameters correspond to the values derived for $r_{\rm h} = 4.5$ and $2.5\,{\rm kpc}$, with the smaller numbers of pressure, magnetic field and density corresponding to the wider hotspot.

\begin{table*}	
	\centering
	\caption{Jet parameters at the hot-spot. The intervals in the parameters correspond to the values derived for $r_{\rm h} = 4.5$ and $2.5\,{\rm kpc}$.}
	\label{tab:parsh}
	\begin{tabular}{cccccccc} 
		\hline
		$\beta_{\rm{h}}$  & $\beta_{\rm{j, h}}$ & $\rho_{\rm{j,h}}$ &$B^{\phi}_{\rm{j,h}}$   &  $P_{\rm h }$ &$B^{\phi}_{\rm h }$ & $\rho_{\rm a}$ \\
		& & [g cm$^{-3}$] &   [mG] &   [dyn/cm$^2$]  &  [mG]  &    [g cm$^{-3}$] \\
		\hline
		0.01 & 0.23 & $1.2 - 3.8 \times 10^{-27}$ & $0.04 - 0.08$ & $0.4 - 1.3 \times 10^{-7}$ & $1.0 - 1.8$  &  $0.5 - 1.5 \times 10^{-24}$\\ 
		0.036 & 0.54 & $0.9 - 2.9 \times 10^{-28}$ & $0.04 - 0.08$ & $1.8 - 5.7 \times 10^{-8}$ & $0.7 - 1.2$  & $1.5 - 5 \times 10^{-26}$\\
		\hline
	\end{tabular}
\end{table*}

We find that the jet velocity at the interaction site is substantially smaller than that of the VLBI jet. This can be explained in terms of kinetic energy dissipation by the growth of the KH modes and/or integrated entrainment along the jet \citep[e.g.,][]{perucho2012a,perucho2012b}. The extremal value of $\beta_{\rm{j, h}} = 0.54$, derived for $\beta_{\rm{h}}=0.036$ provides an a-posteriori justification of our assumption of a mildly relativistic flow because possible relativistics corrections are limited by the Lorentz factor $\Gamma_{\rm j, h} \lesssim 1.2$.

\begin{figure}
	\centering
	\includegraphics[width=\linewidth]{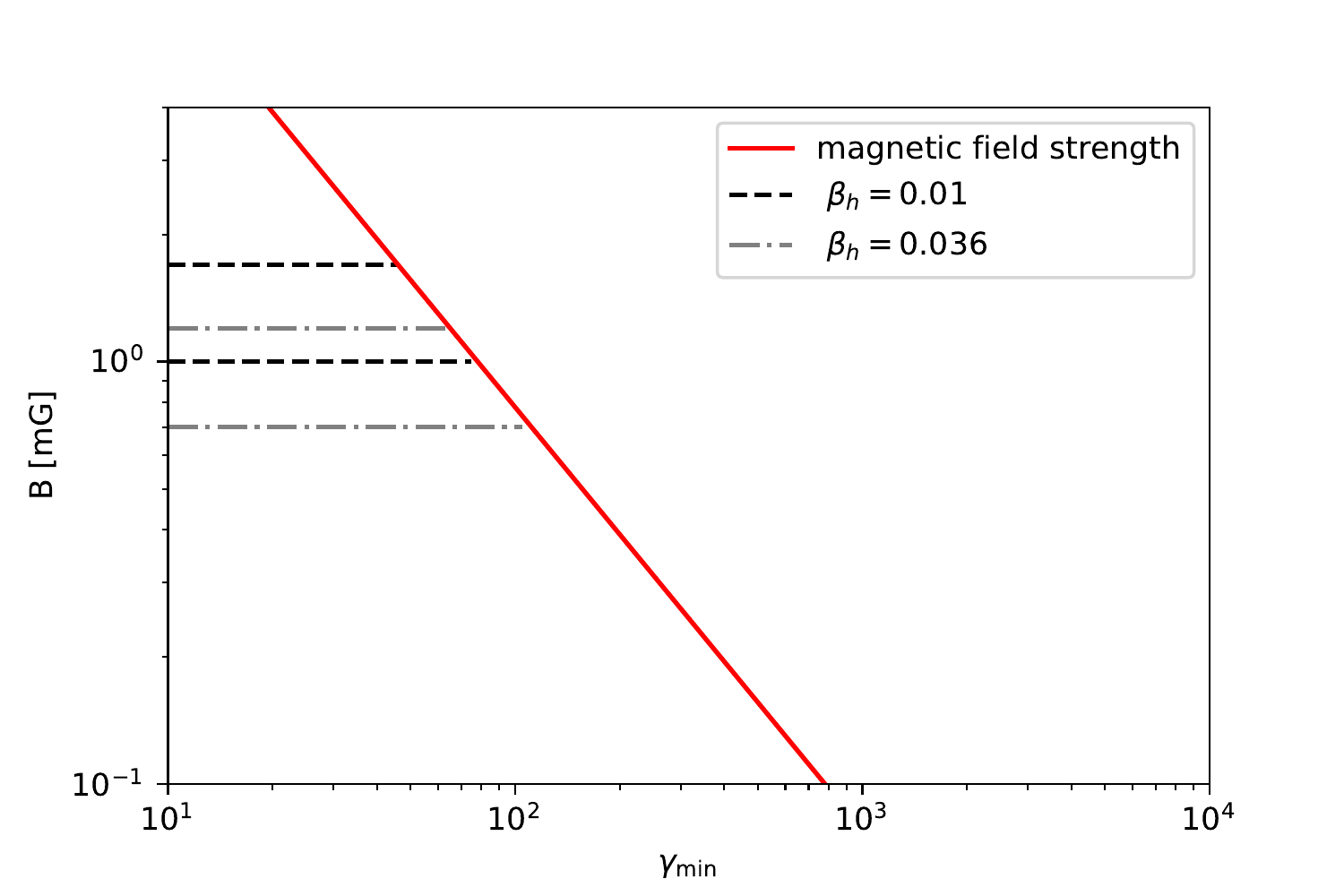}
	\caption{The minimum energy assumption, see \citet{pyrzas2015}, yields magnetic field strengths in the range $0.7\leq \frac{B}{\mathrm{mG}}\leq 1.8$ within the southern hotspot. } 
	
	\label{fig:gammin}
\end{figure}

Our spectral analysis provides us with values for the magnetic field, from basic synchrotron theory, that are displayed in Fig.~\ref{fig:gammin} in terms of the minimum electron Lorentz factor, by using Eq.\,(6) in \citet{pyrzas2015}. The equipartition magnetic field derived for the hotspot region results in a value of $\gamma_{\rm{min}}\leq 100$, which represents a plausible number \citep{meisenheimer1989,meisenheimer1997}.

If we compare the parameters in Table~\ref{tab:parsh} to those obtained by \citet{meisenheimer1989} for different classical FRII hotspots using spectral analysis, we see that our hotspot region values for pressure and magnetic fields are around or slightly above their maximum values ($B_{\rm h} \sim 0.1 - 1$~mG, $P_{\rm h}\sim 0.1 - 1 \times 10^{-8}\,{\rm dyn/cm^2}$ in their case). Taking into account that S5~0836+710 is a powerful jet probably interacting with a dense ICM, we can conclude that the parameters derived by us are in agreement with the typical ones that \citet{meisenheimer1989} obtained using a different method.

From the derived parameters, we can make a step further and estimate the jet density prior to the reverse shock by using Eq.~(\ref{eq:ljc}). The results are also given in Table~\ref{tab:parsh}. The jet number density at the hotspot region lies in the range $n_{\rm j, h} = 0.1 - 4.0\,{\rm cm^{-3}}$ if the mass flux is dominated by pairs or $n_{\rm j, h} = 0.05 - 2.3 \times 10^{-3}\, {\rm cm^{-3}}$ if it is dominated by protons. 

In the case of a leptonic jet, this implies a total particle number flux of $N_{\rm j, e} = \Gamma_{\rm j,\,h} v_{\rm j,\, h} n_{\rm j,\, h} \pi R_{\rm j,\,h}^2 \simeq 1.2 - 5.5 \times 10^{54}\,{\rm pairs/s}$ at the hotspot. This flux falls to $ 0.6 - 3.0 \times 10^{51}\,{\rm s^{-1}}$ in the case of a proton dominated jet at these scales. A pair jet can, nevertheless, be discarded on the basis of energetic argumentation: taking into account that the jet is relativistic at VLBI scales, the energy flux in the form of rest-mass energy must have necessarily changed. As a conclusion, pollution by protons is required to happen along the jet and the jet is likely to be proton dominated on large scales.

\subsection{Implications for the intracluster medium}
\label{sec:intraclust_implications}

From ram pressure confinement, we obtain a minimum value of $\rho_{\rm{a}} = P_{\rm h} /v_{h}^2\simeq 1.5 \times 10^{-26}\,{\rm g\,cm^{-3}}$, and a maximum of $1.5 \times 10^{-24}\,{\rm g\,cm^{-3}}$ for
the possible ranges of $R_j$ and $v_{\rm h}$. The maximum derived (which corresponds to the smallest hotspot advance speed) would imply proton number densities of $\sim 1\,{\rm cm^{-3}}$, of the order of interstellar medium values, and it is therefore unrealistically large for the intracluster medium at 240\,kpc from the active nucleus. Increasing the hotspot advance speed up to $0.036$\,c can yield values as small as $\sim 0.01\,{\rm cm^{-3}}$. The dependencies of the derived ambient density on the observational parameters thus favour hotspot advance speeds  in the upper range of the interval given by brightness asymmetry.  
This result is, however, still one to two orders of magnitude above the values found by \citet{odea2009}. In that work, no clear trend of the density of the intraluster medium is found with redshift out to values of $z \lesssim 1.8$. 
S5~0836+710 is located at $z=2.22$ so that we are in principle probing the density of the intracluster medium at a somewhat earlier evolutionary stage of the expanding universe.
However, the study of a single source does not allow us to draw any conclusions on possible systematic cosmological effects. S5~0836+710 might rather lie within a particular overdense cluster such as the local FR\,II radio galaxy Cygnus\,A. Nevertheless, our method can in principle be applied to large samples of high-power blazars and has the potential to reach out to even higher redshifts.

\section{Conclusion}
\label{sec:conclusion}
The LOFAR telescope provides unprecedented sensitivity and angular resolution in the 100\,MHz regime. For compact sources with structure only on angular scales of arcseconds and smaller, the international LOFAR stations can be used effectively as a VLBI array. Further improvements in image fidelity can be achieved by inclusion of Dutch core-array stations (which can be obtained by improved calibration techniques) and in future observations by the 
new international stations in Poland, Ireland and Italy, yielding a maximum baseline of $\sim 1900$\,km. For the observation of blazars, the unique LOFAR capabilities are crucial because of the small angular scales involved and the beamed core emission, which dominates over the extended structure at higher frequencies. Future LOFAR studies of blazar samples will be able to address unsolved questions about the unification of blazars and radio galaxies, such as the occurance of FR\,II-typical morphological features in BL\,Lac objects \citep{cooper2007,kharb2010}.

In this paper, we have demonstrated that blazar observations with LOFAR can also be used to probe the intracluster medium out to cosmological distances. 
{Our results suggest that the density of the intergalactic medium around the distant ($z=2.22$) blazar S5\,0836+710 might be substantially higher than the values found in less distant FR\,II radio galaxies. However, no generalized statement on a systematic redshift dependence can be derived from a single-source study as presented here, because systematic uncertainties and source-specific peculiarities have to be considered.}
{Our} method can be generalized and applied to larger numbers of suitable blazars to yield statistically relevant samples of the density of the intracluster medium as a function of redshift, which is independent and complementary to classical observational methods applied to radio galaxies \citep[e.g.,][]{odea2009}. Because of their extreme power and beaming, blazars can be found at higher redshift than radio galaxies so that the method might eventually prove particularly important for studies of the young universe. The weakly beamed components of interest (hotspots, lobes), however, will be very challenging to detect at redshifts much larger than $z=2$ and might have to await the advent of the Square Kilometre Array (SKA).


\begin{acknowledgements}
	We thank the anonymous referee for the careful review, which has helped us to improve this paper. This work has been supported by the Spanish Ministerio de Econom\'{\i}a y Competitividad (grants AYA2015-66899-C2-1-P and AYA2016-77237-C3-3-P) and the Generalitat Valenciana (grant PROMETEOII/2014/069).
\end{acknowledgements}   

\appendix 
\section{Light-travel effects on the symmetry between hotspot and counter-hotspot}
\label{sec:appendix_A}

Because the jet in S5~0836+710 is observed at a small angle to the line of sight, we expect a travel time delay between the photons arriving to us from the counter-hotspot and from the hotspot itself. We can estimate the relative observed hotspot to counter-hotspot age and distance to the core by taking into account that the age of the counter-hotspot is $t_{\mathrm{ch}} \simeq D_{\mathrm{ch}} / v_\mathrm{h}$ (see Fig.~\ref{fig:sketch}), assuming that the head of the jet propagates at a similar speed for long distances.
Furthermore, we can also claim that $t_{\mathrm{ch}} = t_{\mathrm{h}} - D/c$, i.e., the counter-hotspot is seen at a younger age than the hotspot, with the difference due to the time the photons need to travel the total distance between the two hotspots, $D=D_\mathrm{h}+D_{\mathrm{ch}}$. By combining these expressions, we get:
\begin{equation}
\frac{D_{\mathrm{ch}}}{v_\mathrm{h}} = t_\mathrm{h} - \frac{(D_{\mathrm{h}}+D_{\mathrm{ch}})}{c},
\end{equation}
and thus:
\begin{equation}
D_{\mathrm{ch}} = \left(\frac{D_\mathrm{h}}{v_\mathrm{h}}-\frac{D_{\mathrm{h}}}{c}\right)\,\frac{1}{\left(\frac{1}{v_\mathrm{h}}+\frac{1}{c}\right)},
\end{equation}
which, for $v_\mathrm{h}=0.023\,c$ and $D_{\mathrm{h}} \simeq 240\,{\rm kpc}$ ($\vartheta = 3^\circ$), result in $D_{\mathrm{ch}} = 229\,{\rm kpc}$, and an age of about 5 Million years younger then the southern hotspot.

\section{Jet power}
\label{sec:appendix_B}

In the case of a non-relativistic jet, Eq.~\ref{eq:ljr} becomes
\begin{equation}\label{eq:ljc}
L_{\rm j}\,=\, \left( \frac{\gamma_{\rm j}\,P_{\rm j}}{(\gamma_{\rm j}-1)}
\,+\, \frac{1}{2} \rho_{\rm j} v_{\rm j}^2 \,+\,\frac{(B^{\phi})^2}{4\pi} \right) \, \, v_{\rm j}\,A_{\rm j}.
\end{equation}

\cite{VegaGarcia2018phd} has provided an estimate of the mean magnetic field along the 1.6~GHz jet of S5\,0836+710 by using the cooling time of the electrons emitting at this frequency together with the distance along which the jet is observed, resulting in $B\lesssim 0.01\,{\rm G}$. This represents an upper limit of the field strength, as it obviates adiabatic cooling on the scales observed. Using this value, the jet radius at half the jet extension, and the estimated Lorentz factor for the VLBI jet ($\Gamma_{\rm j}=12$), \cite{VegaGarcia2018phd} derive a jet Poynting flux $L_p = \frac{(B^{\phi})^2}{4\pi} v_{\rm j} \pi R_{\rm j}^2 \simeq 2.8\times10^{45}\,{\rm erg/s}$, assuming that the dominating field is toroidal.

Following an independent line of derivation, we use the intrinsic hotspot plus counter hotspot luminosities at 150\,MHz ($9.0\times 10^{29}\,{\rm W\,Hz^{-1}}$, altogether) and the relation given by \citet{daly2012} between the source luminosity at 178~MHz and the jet kinetic power to estimate the latter, and we obtain $L_{\rm j}\simeq 10^{47}\,{\rm erg/s}$. Although this value is affected by considerable uncertainties, it yields a number corresponding to that expected for powerful FR\,II jets, as expected for this bright and distant quasar. By comparing the total energy flux with the Poynting flux, we can deduce that the jet is likely to be kinetically dominated at these scales. 

\begin{table*}	
	\centering
	\caption{Jet parameters at VLBI scales.}
	\label{tab:parsj}
	\begin{tabular}{cccccc} 
		\hline
		$B^{\phi}_{\rm 1.6\,GHz}$ & $\beta_{\rm j, 1.6\,GHz}$ & $R_{\rm j, 1.6\,GHz}$ & $L_{178\,{\rm MHz}}$ & $L_p$ & $L_{\rm j}$ \\
		$~$[mG]& & [pc] &  [$10^{36}$ erg  s$^{-1}$ Hz$^{-1}$] & [$10^{45}$ erg s$^{-1}]$ &  $[10^{45}$ erg s$^{-1}]$   \\
		\hline
		10 & 0.9965 & 20 & 1.7 & 2.8 & 100\\ 
		\hline
	\end{tabular}
\end{table*}

Stability analysis of this jet confirms, in addition, that its Mach number is large or, in other words, that the jet is in the cold regime ($h_{\rm j} \simeq c^2$ and implying $\gamma_{\rm j}\simeq 5/3$, Vega-Garc\'{\i}a et al., submitted), albeit kinetically relativistic at VLBI scales. Table~\ref{tab:parsj} contains a summary of the jet parameters that are relevant to our study \citep[see][]{VegaGarcia2018phd}.

\section{The relativistic case}
\label{sec:appendix_C}
In Sect.~\ref{sec:jet_param_deriv}, we used the approximation of a sub-relativistic jet flow close to the hot-spot, which proved to be valid because of the limiting flow velocity obtained ($\beta_{\rm j,h} \leq 0.54$). For reference, we also considered the case of a relativistic flow, where the conservation of momentum flux across the shock implies \citep[e.g.,][]{ll59,mm94}: 

\begin{equation} \label{eq:ph}
\frac{\rho_{\rm j} \, h_{\rm j} \, {\bf \Gamma}_{\rm j}^2 \, v_{\rm j}^2}{c^2}+\, P_{\rm j}\, = 
\, \frac{\rho_{\rm h} \, h_{\rm h} \, {\bf \Gamma}_{\rm h}^2 \, v_{\rm h}^2}{c^2} + \, P_{\rm h}.
\end{equation}

For the particle dominated (cold) jet of S5\,0836+710 at these scales (see Appendix~\ref{sec:appendix_B}),  $P_{\rm j}$ is much smaller than $\rho_{\rm j} c^2$ (implying that $h_{\rm j} \simeq c^2$). Taking into account that for a strong shock
$v_{\rm h} \ll v_{\rm j}$, 
it follows that $P_{\rm h} \gg P_{\rm j}$ and the previous expression results in

\begin{equation}\label{eq:p_rel}
P_{\rm h}\simeq \frac{\rho_{\rm j} {\bf \Gamma}_{\rm j}^2 v_{\rm j}^2}{c^2}.
\end{equation}

Comparing Eqs.~(\ref{eq:ljr}) and (\ref{eq:p_rel}), subtracting the rest mass energy of the particles, and neglecting the magnetic field contribution, we obtain, for the hotspot pressure:

\begin{equation}\label{eq:p_Lrel}
P_{\rm h} \simeq \frac{{\bf \Gamma}_{\rm j} L_{\rm j} \, \beta_{\rm j, h}}{({\bf \Gamma}_{\rm j} - 1) \pi R_{\rm j,h}^2 \, c}.
\end{equation}

We approximate the jet radius at the hotspot, $R_{\rm j,h}$, to the hotspot radius.
As we did before, we now begin an iterative process: we start with $\beta_{j,h} \simeq 1$ to find a first value for the hotspot pressure $P_{\rm h}$. Equipartition between the non-thermal particles and the magnetic field can then be used to derive the hotspot magnetic field, $B_{\rm h} \simeq \sqrt{8\pi P_{\rm h}}$. We note that the values derived from Eq.~\ref{eq:p_Lrel} lead to upper limits of the hotspot pressure and magnetic field because there is no total transfer of kinetic energy into internal energy across the shock. This is caused by the approximation of the first term in the right hand side of Eq.~\ref{eq:ph}, $\simeq \rho_{\rm h} \, v_{\rm h}^2$, to zero.

Then, with the value of the magnetic field derived for the jet at VLBI scales and applying conservation of the magnetic flux (which is a reasonable assumption for conical jets), we derive an estimate of the field intensity right before the hotspot. Finally, we can apply the RMHD jump conditions at the reverse shock giving rise to the hotspot \citep[e.g.,][]{anile1989}:
\begin{equation}
\Gamma_{\rm j, h} \beta_{\rm j, h} B^{\phi}_{\rm j, h} \, = \, \Gamma_{\rm h} \beta_{\rm h} B^{\phi}_{\rm h},
\end{equation}   
which, using the value derived for $\beta_{\rm{h}}$, $\Gamma_{\rm{h}} \simeq 1$, $B^{\phi}_{\rm{h}}$, and $B^{\phi}_{\rm{j, h}}$, allows us to obtain  $\Gamma_{\rm{j, h}}~ \beta_{\rm j,h}$. From the value of $\Gamma_{\rm{j, h}}~ \beta_{\rm j,h}$, we get $\beta_{\rm{j, h}}$ and bring it back to Eq.~\ref{eq:p_Lrel} to continue the iterative method.

In the case of S5\,0836+710 with its very low measured head velocity $\beta_{\rm h}$, the relativistic expression necessarily leads to the classical approximation, because the kinetic energy, $(\Gamma_{\rm j,h} - 1) \rho_{\rm j,h} c^2 v_{\rm j,h} A_{\rm j,h}$ tends to zero. The Taylor expansion of the Lorentz factor results in the classical kinetic energy.

\bibliographystyle{aa}


\end{document}